\documentclass[fleqn,usenatbib]{mnras}

\usepackage{newtxtext,newtxmath}

\usepackage[T1]{fontenc}

\DeclareRobustCommand{\VAN}[3]{#2}
\let\VANthebibliography\thebibliography
\def\thebibliography{\DeclareRobustCommand{\VAN}[3]{##3}\VANthebibliography}

\usepackage{graphicx}	
\usepackage{amsmath}	
\usepackage{booktabs}
\usepackage[dvipsnames]{xcolor}
\usepackage{ulem}

\newcommand{\vlos}{v_\mathrm{los}}
\newcommand{\xproj}{x_\mathrm{proj}}
\newcommand{\yproj}{y_\mathrm{proj}}
\newcommand{\Rproj}{R_\mathrm{proj}}

\newcommand{\mpc}{\operatorname{Mpc}}
\newcommand{\msun}{\mathrm{M}_\odot}
\newcommand{\hmsun}{h^{-1}\msun}
\newcommand{\hmpc}{h^{-1}\mathrm{Mpc}}
\newcommand{\kms}{\mathrm{km}\ \mathrm{s}^{-1}}
\newcommand{\mthc}{M_\mathrm{200c}}

\newcommand{\mfhc}{M_\mathrm{500c}}
\newcommand{\rfhc}{R_\mathrm{500c}}

\newcommand{\lx}{L_X}
\newcommand{\mgas}{M_\mathrm{gas}}
\newcommand{\mstar}{M_\mathrm{star}}
\newcommand{\sigvtrue}{\sigma_\mathrm{v,1D,true}}
\newcommand{\sigv}{\sigma_\mathrm{v,1D}}
\newcommand{\ngal}{N_\mathrm{gal}}
\newcommand{\nphot}{N_\mathrm{phot}}
\newcommand{\nphotfhc}{N_\mathrm{phot,500c}}

\newcommand{\zclu}{z_\mathrm{clu}}

\newcommand{\mtrue}{M_\mathrm{true}}
\newcommand{\mpred}{M_\mathrm{pred}}

\newcommand{\rev}[1]{{#1}}

\title[Deep Learning X-ray Masses of Galaxy Clusters]{Benchmarks and Explanations for Deep Learning Estimates of X-ray Galaxy Cluster Masses}

\author[M. Ho et al.]{
Matthew Ho,$^{1}$\thanks{E-mail: matthew.annam.ho@gmail.com}
 John Soltis,$^{2}$
 Arya Farahi,$^{3}$
 Daisuke Nagai,$^{4}$
 August Evrard,$^{5}$
 and Michelle Ntampaka$^{2,6}$
\\
$^{1}$CNRS \& Sorbonne Universit\'{e}, Institut d’Astrophysique de Paris (IAP),
UMR 7095, 98 bis bd Arago, F-75014 Paris, France\\
$^{2}$Department of Physics \& Astronomy, Johns Hopkins University, Baltimore, MD 21218, USA\\
$^{3}$Departments of Statistics and Data Science, University of Texas at Austin, Austin, TX 78705, USA\\
$^{4}$Department of Physics, Yale University, New Haven, CT 06520, USA\\
$^{5}$Departments of Physics and Astronomy and Leinweber Center for Theoretical Physics, University of Michigan, Ann Arbor, MI 48109, USA\\
$^{6}$Data Science Mission Office, Space Telescope Science Institute, Baltimore, MD, 21218, USA
}

\date{Accepted XXX. Received YYY; in original form ZZZ}

\pubyear{2023}

\begin{document}
\label{firstpage}
\pagerange{\pageref{firstpage}--\pageref{lastpage}}
\maketitle

\begin{abstract}
We evaluate the effectiveness of deep learning (DL) models for reconstructing the masses of galaxy clusters using X-ray photometry data from next-generation surveys. We establish these constraints using a catalogue of realistic mock eROSITA X-ray observations which use hydrodynamical simulations to model realistic cluster morphology, background emission, telescope response, and AGN sources. Using bolometric X-ray photon maps as input, DL models achieve a predictive mass scatter of \rev{$\sigma_{\ln\mfhc} = 17.8\%$}, a factor of two improvements on scalar observables such as richness $\ngal$, 1D velocity dispersion $\sigv$, and photon count $\nphot$ as well as a \rev{$32\%$} improvement upon idealised, volume-integrated measurements of the bolometric X-ray luminosity $\lx$. We then show that extending this model to handle multichannel X-ray photon maps, separated in low, medium, and high energy bands, further reduces the mass scatter to \rev{$16.2\%$}. We also tested a multimodal DL model incorporating both dynamical and X-ray cluster probes and achieved marginal gains at a mass scatter of \rev{$15.9\%$}. Finally, we conduct a quantitative interpretability study of our DL models and find that they greatly down-weight the importance of pixels in the centres of clusters and at the location of AGN sources, validating previous claims of DL modelling improvements and suggesting practical and theoretical benefits for using DL in X-ray mass inference.
\end{abstract}




\section{Introduction}

Galaxy clusters are large collections of dark matter, gas, and galaxies dynamically bound through gravitational attraction.
They are the most massive systems in the Universe, and, as such, act as observable tracers of peak density regions in the cosmic web. The spatial and temporal distribution of galaxy clusters encodes information regarding the growth of large-scale structures throughout the Universe's evolution \citep[][for a review]{allen2011cosmological}. For example, the abundance of clusters as a function of mass is a widely used probe of the Universal matter density $\Omega_m$ and the amplitude of primordial density fluctuations $\sigma_8$ \citep[e.g.,][]{Vikhlinin2009, Mantz2010, ade2016planck, abbott2020dark}. A plethora of multiwavelength cluster surveys are underway to constrain physical models of gravity, dark matter, and dark energy \citep[e.g.,][]{pillepich2018forecasts,Raghunathan2022}. 

Accurate and precise estimates of galaxy cluster masses are of paramount importance to these analyses, as a cluster's mass sets the scale for processes both within the system and in its interactions with the cosmic web \citep[][for a recent review]{pratt2019galaxy}. \rev{However, 90\% of a cluster's mass is hidden away in its host dark matter halo and must be inferred  indirectly from the spatial and energy distributions of photons observed on the sky.}
A wide diversity of techniques has been developed to constrain cluster masses using data from X-ray \citep[e.g.,][]{kravtsov2006new,Pratt2009,2017MNRAS.465..858G,Mantz2016WtG}, microwave \citep[e.g.,][]{Nagai2006,Kay2012}, and optical surveys \citep[e.g.,][]{saro2013toward, wojtak2018galaxy}.
Generally, these methods establish a physical connection between observable signals and cluster masses and then use simulations \citep[e.g.,][]{nagai2007testing, planelles2014role, biffi2016nature, Pop2022a, Pop2022b} or complementary observations \citep[e.g.,][]{ade2011planck, reichardt2013galaxy, schellenberger2015xmm, Mulroy2019, mcclintock2019dark} to calibrate their mass inference. 

The recent popularity of this problem has led to considerable work to reduce the level of mass scatter with novel data analysis methods. \rev{These approaches benefit from their ability to utilise complex signals in cluster observables for mass inference, which are otherwise difficult to model analytically.} Deep neural networks trained on mock observations from hydrodynamical simulations have shown a strong ability to improve the accuracy of cluster mass estimates using X-ray \citep{ntampaka2019deep,Green2019,yan2020galaxy,krippendorf2023erosita}, microwave \citep{Cohn2019,wadekar2022augmenting, wadekar2022sz, de2022deep}, and optical data \citep{ntampaka2015,ho2019robust, ho2021approximate, ho2022dynamical, kodi2020dynamical}. These techniques, while promising, require extensive validation on mock data before they can be reliably extended to observational samples \citep{ntampaka2021building}.

Today, the development of techniques for precise mass inference on X-ray images is of particular significance due to the recent launch of the extended ROentgen Survey with an Imaging Telescope Array \citep[eROSITA][]{merloni2012erosita}. eROSITA is designed to perform a deep survey of the sky through the X-ray energy band ($0.5-10\ \mathrm{keV}$). It is projected to detect approximately $\sim100,000$ clusters over its four-year operating timeline \citep{pillepich2018forecasts}. eROSITA will observe these clusters at a lower angular resolution than its predecessor, the \textit{Chandra} X-ray Observatory, but will reach a much larger sample at a well-modelled selection function, making it an ideal instrument for cosmological inference. In addition, the clusters detected by eROSITA will be subject to follow-up with spectroscopic, optical, and microwave instruments of related surveys, such as SDSS-V, DESI, Euclid, and Rubin, opening the door toward large-scale multiwavelength studies of cluster physics. The use of accurate, precise, and efficient techniques for inference is essential to capitalise on the wealth of eROSITA data.

In this paper, we forecast and explain the mass estimation performance of neural network analysis of X-ray observations for the recently launched eROSITA telescope. We utilize a catalogue of mock observations of $3,285$ distinct clusters in the Magneticum hydrodynamical simulation \citep{dolag2016sz}, each including realistic noise, instrument response, and simulation-driven realisations of cluster morphology, core physics, and AGN contamination. We train and test modern neural network models on this catalogue and benchmark their mass predictions against the common observable mass proxies. We implement methods for further reducing scatter using multiwavelength probes, specifically by applying neural networks on multi-band X-ray images and joint X-ray and spectroscopic data. We then conduct a quantitative interpretability study of our ML models, investigating and evaluating the improved behaviour of neural networks in the context of cluster mass prediction.

The structure of this paper is as follows. Section~\ref{sec:dataset} presents the simulation data and the procedure for generating mock observations. Section~\ref{sec:baseline} describes our comparative baseline, a covariance analysis of scalar mass proxies commonly used in X-ray analysis. Section~\ref{sec:xray_only} introduces our CNN modelling approach for single-band X-ray images and reports their predictive performance against our baselines. Section~\ref{sec:multiwavelength} describes the performance improvements of CNN models, which incorporate multiband X-ray and spectroscopic dynamical information. Section~\ref{sec:discussion} explains the results of our interpretability study of the aforementioned CNN models. Lastly, we present our conclusions and suggestions for future work in Section \ref{sec:conclusion}. 

\section{Data}\label{sec:dataset}

\begin{table}
    \centering
    \begin{tabular}{c|c|c|c}
        \toprule
        Property & Min. & Median & Max.\\
        \midrule
        $\log_{10}\left[\mfhc\ (\hmsun)\right]$ & $13.50$ & $13.94$ & $15.07$\\
        $z$ & $0.07$ & $0.21$ & $0.47$\\
        $\nphot$ & $17$ & $1072$ & $2.37\times 10^{5}$\\
        $N_\mathrm{ICM}$ & $9$ & $762$ & $2.00\times 10^{5}$\\
        $N_\mathrm{AGN}$ & $0$ & $225$ & $2.03\times 10^{5}$\\
        $\ngal$ & $28$ & $112$ & $520$\\
        \bottomrule
    \end{tabular}
    \caption{Summary properties of the mock X-ray (Section \ref{subsec:data_xray}) and dynamical catalogues (Section \ref{subsec:data_spec}). Each row shows the minimum, median, and maximum values of all clusters in our training and test catalogues. The quantities, in order, are: the logarithmic cluster mass, redshift, total observed X-ray photons, total observed photons from ICM sources, total observed photons from AGN sources, and the number of galaxies within the dynamical selection cut (Section \ref{sec:baseline}, for details). The photon counts reported here are the values prior to the imposed redshift normalisation with Equation \ref{eqn:znorm}.}
    \label{tab:dataset}
\end{table}

\begin{figure*}
    \centering
    \includegraphics[width=0.95\linewidth]{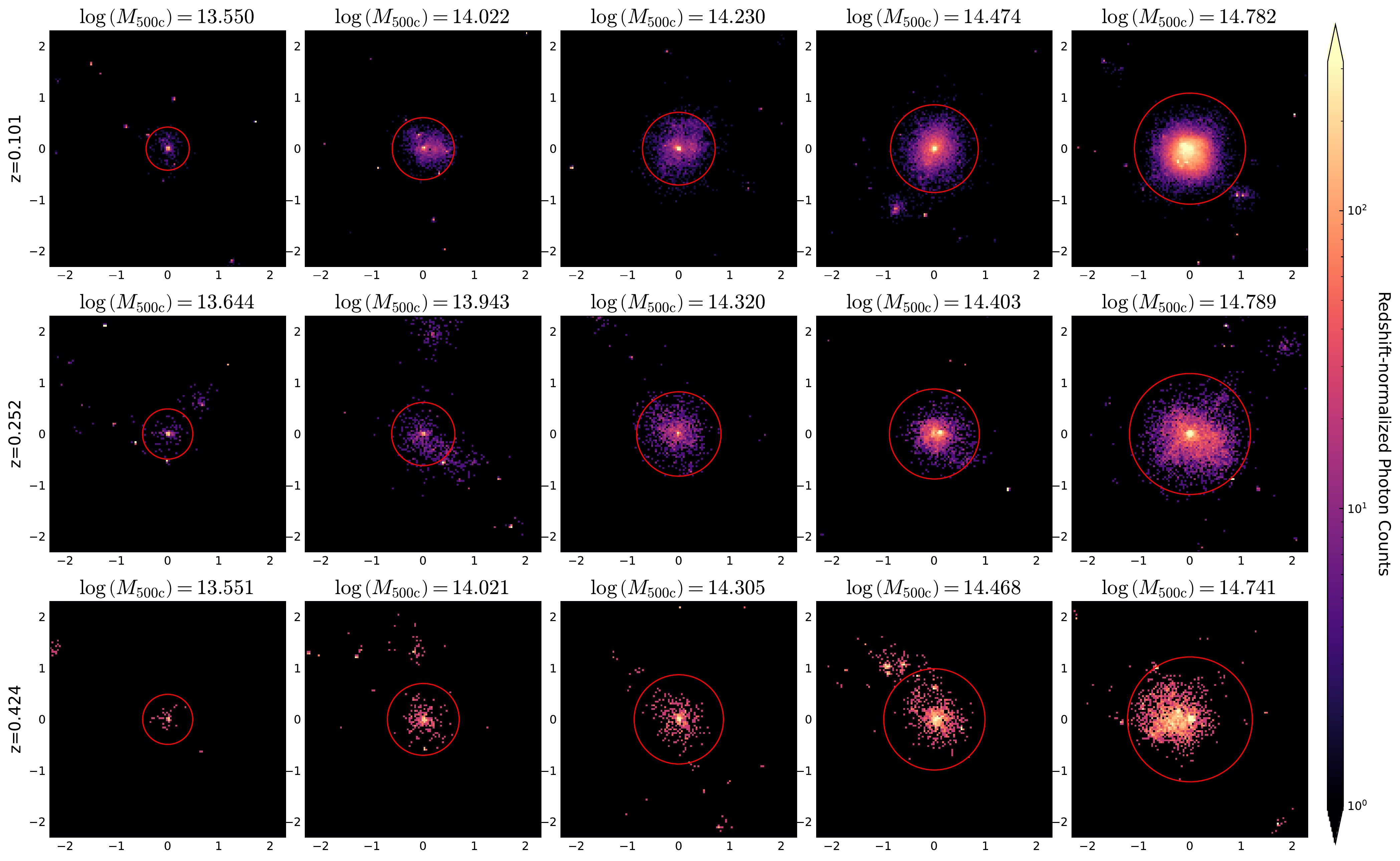}
    \caption{Mock eROSITA photon maps for twelve example clusters in the Magneticum simulation. Each row selects clusters from a different Magneticum snapshot, and selected clusters are sorted in increasing mass from left to right. Photon counts in each sky-projected bin are normalised in scale to account for the system redshift (Equation \ref{eqn:znorm}). Each subplot also shows a red circle indicating the $\rfhc$ of each cluster. $\mfhc$ masses are presented in units of $\hmsun$ and subplot x- and y-axes are labelled in units of $\hmpc$.}
    \label{fig:examples}
\end{figure*}

Successful implementation of data-driven inference methods, such as the neural networks described in Section~\ref{sec:xray_only} and Section~\ref{sec:multiwavelength}, requires training catalogues that accurately reflect the realism and distribution of observational data. The mock catalogues curated in this analysis are among the most realistic X-ray cluster mocks currently applied to deep learning models and are the first to incorporate realistic modelling of AGN sources \citep[See][for a comparison]{ntampaka2019deep,Green2019,yan2020galaxy}. Our dataset is built on mock eROSITA catalogues generated in \citet{soltis2022machine}. In the following sections, we describe the primary properties of the mock catalogue, as well as adjustments made to it for the purposes of this work. For details on the original catalogue, see \citet{soltis2022machine}.

The base simulation for our mock catalogue is the Magneticum Box 2/hr hydrodynamical simulation \citep{hirschmann2014cosmological}, provided by the Cosmological Web Portal \citep{ragagnin2017web}. The Magneticum simulation \citep{dolag2016sz} is a cosmological TreePM-SPH simulation code containing physical models for radiative cooling, star formation, stellar populations, and black hole and AGN feedback. Hydrodynamical simulations such as Magneticum are a key cornerstone of our understanding of galaxy clusters and have been used extensively to study X-ray properties of clusters \citep[e.g.,][]{kravtsov2006new, nagai2007, biffi2018agn, Green2019}. 
 The Box2/hr run is simulated in a cube of side length $352\ \hmpc$, an initial particle number of $2\times 1564^3$, a dark matter particle mass of $M_\mathrm{DM} = 6.9\times 10^{18}\ M_\odot$, and a gas mass of $M_\mathrm{gas}=1.4\times 10^8\ M_\odot$. It assumes a WMAP7 cosmology of $\Omega_m = 0.272$, $\Omega_b = 0.0456$, $\Omega_\Lambda = 0.728$, $h=0.704$, $n_s = 0.963$, and $\sigma_8 = 0.809$ \citep{komatsu2009five}. The halos and subhalos in this volume are identified via the Subfind Friends-of-Friends halo finder \citep{springel2001populating, dolag2009substructures}. Throughout this work, halo mass is defined as the spherical overdensity mass at a density contrast of 500 times the critical density of the universe, henceforth referred to as $\mfhc$ in the cosmology invariant units of $\hmsun$.

We selected a sample of 3285 galaxy clusters from the Magneticum halo catalogue, chosen to have a roughly uniform distribution across the redshift range $0.07 \leq z \leq 0.47$ and across the mass range $3.16\times 10^{13} M_\odot \leq \mfhc \leq 1.17\times 10^{15} M_\odot$.  This was done to train our inference model to have a uniform prior over mass and redshift, a favourable property for cosmological analyses. The distribution of the cluster properties of our sample is characterised by Table \ref{tab:dataset}. Each simulated cluster is viewed from a single projected perspective along the $z$ box axis, and is therefore included in our mock catalogue only once. In the following subsections, we describe the X-ray and spectroscopic observables calculated for each mock cluster from this projected perspective.


\subsection{Mock eROSITA Observations} \label{subsec:data_xray}

\rev{The mock X-ray observations used in this work are designed to match the observing conditions} described in the eROSITA instrument specifications \citep{merloni2012erosita}. Specifically, our observations project a field of view spanning approximately $1^{\circ}$, a pixel size of $9.6\arcsec$, and an observation time of $2\ \mathrm{ks}$. The centre of each cluster is assumed to be located at a sky position of $10\arcmin$ right ascension and $10\arcmin$ declination.  We have simulated intra-cluster medium (ICM) emission using the PHOX algorithm \citep{biffi2012observing, biffi2013investigating} incorporating an observation depth of $10\ \mpc$. Light from AGN sources simulated in the Magneticum volume is projected into our observations using the method described in \citet{biffi2018agn}. \rev{These sources trace the positions and properties of Stellar Mass Black Holes (SMBH) in the Magneticum simulation and thus follow the distribution of large-scale structure contaminants expected in real observations.} Additionally, we have generated realisations of background emission, instrument response, and point spread function that are consistent with the eROSITA telescope design and exposure using the SIXTE algorithm \citep{dauser2019sixte}.  For more information on modelling X-ray contamination in the eROSITA telescope with SIXTE, see \citet{clerc2018synthetic}.

The SIXTE software produces a list of photons expected to be observed by eROSITA for each simulated cluster. Table \ref{tab:dataset} tabulates the minimum, median, and maximum number of photons that we receive in these images, from ICM and AGN sources. \rev{Photons observed within each mock SIXTE pixel are assumed to sit at their pixel center.} We then map these photon \rev{sky coordinates to flat-sky 2D images}. First, we convert the photon list in projected sky coordinates to comoving coordinates using the Magneticum WMAP7 cosmology \citep{komatsu2009five}. We then measure a projected 2D histogram of photon counts across a square aperture of length $2.3\ \hmpc$ at a pixel resolution of $128\times 128$. 
\rev{This has the effect of `zooming-in' high-redshift systems and `zooming-out' low-redshift systems, resulting in a set of 2D photon maps which are scale-invariant to redshift.. Since this data is derived from the same angular resolution and instrumental effects as the SIXTE configuration, this operation is repeatable on real observations.} We also normalise the photon counts in each pixel to luminosity distance by assuming a log-linear relationship between redshift and total measured photons. Explicitly, the luminosity normalisation follows
\begin{equation}\label{eqn:znorm}
    n_{ij} = m_{ij}\times\frac{N_0}{10^{\gamma\zclu + \eta}},
\end{equation}
where $m_{ij}$ and $n_{ij}$ are the raw and redshift-normalised per-pixel photon counts, respectively, $N_0$ is a pivot value fixed at an approximate mean total photon count $4000$, $\zclu$ is the redshift of the cluster, and $m$ and $b$ are linear regression slopes learned from fitting $\log \sum_{ij}m_{ij} = \gamma\zclu + \eta$ for all mock observations in our training set.
These normalising transformations produce a set of images whose angular scale and absolute magnitude are invariant to redshift, apart from the shot noise and AGN contamination associated with their evolved environments. We note here that these operations are possible with knowledge of the cluster redshift, which will be attainable for eROSITA clusters through the SPIDERS spectroscopic followup program \citep{clerc2020spiders}.

From this step, we derive two versions of the X-ray images. We first produce single-band images integrated over the full energy range of the eROSITA instrument ($0.5-10\ \mathrm{keV}$). These are to be used in our CNN modelling of single-band photon maps, as presented in Section~\ref{sec:xray_only}. Examples of these images are shown in Figure~\ref{fig:examples}. We also produce multi-band images in the three eROSITA energy bands, namely the soft ($0.5-1.2\ \mathrm{keV}$), medium ($1.2-2.0\ \mathrm{keV}$), and hard bands ($2.0-7.0\ \mathrm{keV}$). These are to be used in Section~\ref{sec:multiwavelength} as an improved X-ray probe of the cluster mass. The sole difference between the multi-band and single-band images is in their energy separation, i.e., the total photons received by each pixel across all energies in the multi-band are equal to those of the single band. An example of a multi-band image is shown in Figure \ref{fig:ex_multi}.

Lastly, to reduce the dynamic range of the pixel values, we apply a logarithmic scaling to the images before they are passed as input to the neural network. The scaling we use is,
\begin{equation}\label{eqn:inputscaling}
    x_{ij} = \log_{10}\left[n_{ij}+1\right],
\end{equation}
where $n_{ij}$ is the redshift-normalised number of photons detected at the $j$-th pixel of the $i$-th row, and $x_{ij}$ is the corresponding scaled pixel value used as input of the neural network.

\subsection{Mock Spectroscopic Follow-up}\label{subsec:data_spec}

In addition to the X-ray images, we generated a matching catalogue of spectroscopic follow-up observations to evaluate the multiwavelength models in Section \ref{sec:multiwavelength}. These catalogues follow the design of those presented in \citet{ho2019robust, ho2021approximate}. We assume that an observer at redshift $z=0$ measures the exact spectra of galaxies around each Magneticum cluster from the same \rev{line-of-sight} perspective as our X-ray observations. The galaxy positions and velocities are seeded from Magneticum subhalos found with the Subfind algorithm. We then convert these measurements to comoving positions, velocities, and stellar mass estimates. We place cuts on these measured properties to assign cluster membership to galaxies with stellar mass $M_\mathrm{star} \geq 10^{9.5}\ \hmsun$ and within a dynamical cylinder of projected radius $\Rproj \leq 2.3 \hmpc$ and line-of-sight velocity $|\vlos|\leq 3785\ \kms$, relative to the cluster center. With this large dynamical cut, we incorporate the various systematics that impact observational studies of cluster dynamics, including simulated physics of dynamical substructure \citep{old2018galaxy}, cluster mergers \citep{Evrard2008}, halo environment \citep{white2010cluster}, and interloping galaxies \citep{wojtak2018galaxy}. Table \ref{tab:dataset} shows the minimum, median, and maximum number of galaxies in each observation after selection cuts. 

\begin{figure}
    \centering
    \includegraphics[width=\linewidth]{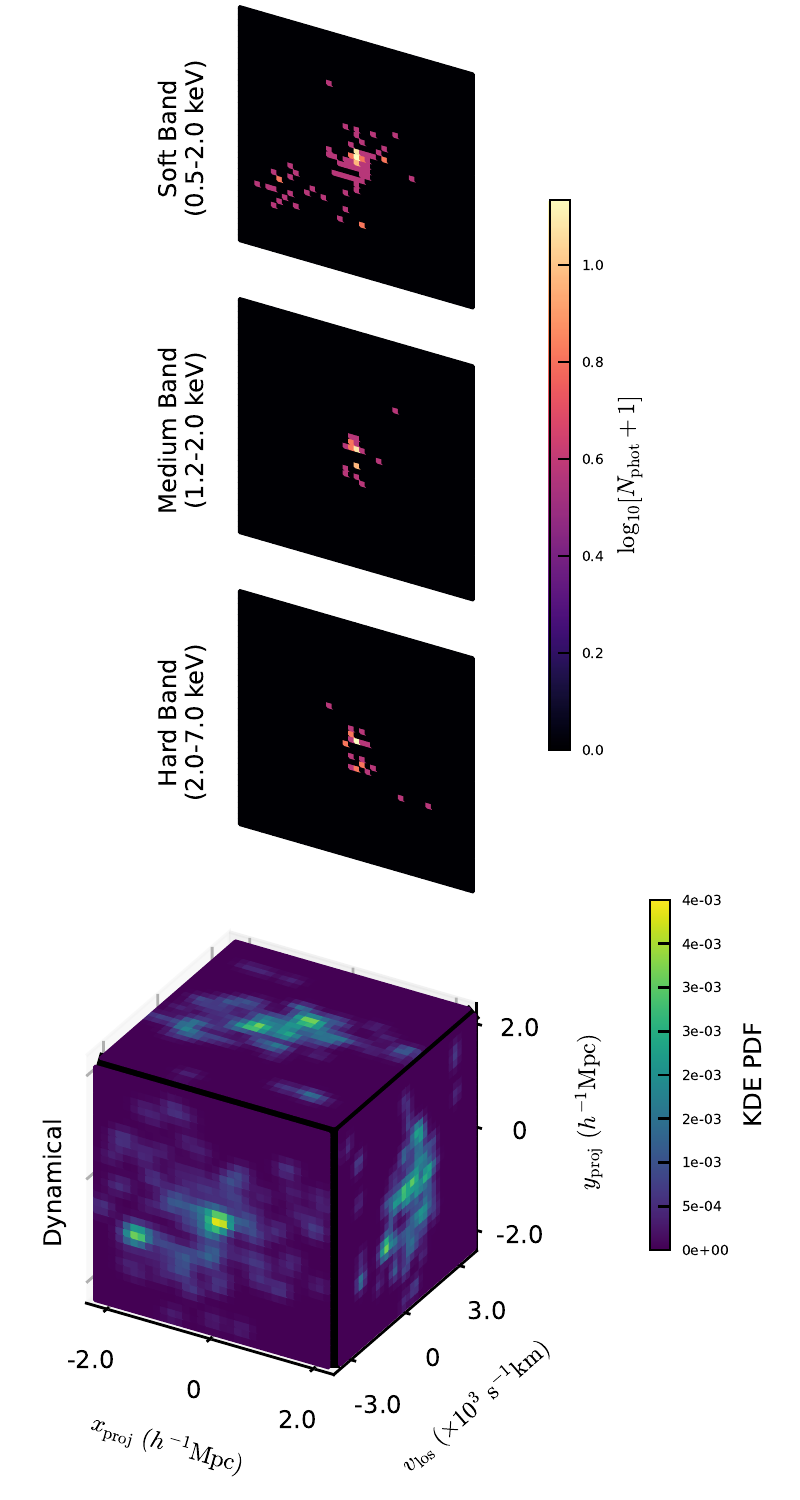}
    \caption{Example multiwavelength observables for a single Magneticum cluster at redshift $z=0.47$ and mass $\mfhc = 1.7\times 10^{14}\ \hmsun$. The top three images show mock eROSITA photon maps in the soft, medium, and hard bands. The bottom subplot shows the KDE-estimated distribution of galaxies in dynamical phase space $\{\xproj, \yproj, \vlos\}$. The X-ray images and the dynamical cube are oriented identically such that the line of sight is oriented to the top-right, as indicated by the red arrow.}
    \label{fig:ex_multi}
\end{figure}

To turn these galaxy catalogues into regular inputs amenable to CNN architectures, we apply the density estimation and sampling technique introduced in \citet{ho2019robust} and extended in \cite{kodi2021simulation}. For each cluster in our training set, we use kernel density estimators \citep[KDEs;][]{scott2015multivariate} to estimate the distribution of galaxies in the 3D dynamical phase space, $\{\xproj, \yproj, \vlos\|$. The KDEs applied here are Gaussian with a fixed bandwidth scaling factor of $0.15$. We then sample this distribution on a regular $64\times64\times64$ grid that spans our dynamic selection cut, that is, $|\vlos|\leq 3785\ \kms$, $|\xproj| \leq 2.3 \hmpc$, and $|\yproj| \leq 2.3 \hmpc$. This operation has the effect of smoothing our discrete list of galaxy positions and velocities into a fixed-size input. The resulting 3D input box is crucial to utilising convolutional filters in our neural network. Figure \ref{fig:ex_multi} shows an example of the distribution of galaxies around a Magneticum cluster in a cube of dynamical phase space.

\section{Baseline Scalar Observables}\label{sec:baseline}

\begin{table}
    \centering
    \begin{tabular}{c|l}
        \toprule
        Observable & Description \\
        \midrule
        $T$ & Temperature \\
        $\lx$ & Bolometric X-ray luminosity \\
        $\mgas$ & Gas mass within $\rfhc$\\
        $\mstar$ & Stellar mass within $\rfhc$\\
        $\sigvtrue$ & Projected 1D velocity dispersion\\
        \midrule
        $\ngal$ & Richness \\
        $\sigv$ & Total projected 1D velocity dispersion\\
        $\nphot$ & Total received photon count\\
        $\nphotfhc$ & Photon count within $\rfhc$\\
        \bottomrule
    \end{tabular}
    \caption{Baseline scalar observable proxies for cluster mass. The top five quantities represent `idealised' measurements of these observables, while the bottom four are `realistic' measurements contaminated with observational systematics. Cluster centres, idealised hydrodynamical properties, and $\rfhc$'s are produced by the \textsc{Subfind} algorithm \citep{dolag2009substructures} and reported in \citet{ragagnin2017web}.}
    \label{tab:observables}
\end{table}

We first develop a quantitative baseline for predicting the host halo mass using scalar observables. Table~\ref{tab:observables} details the complete list of observables used in this analysis. A subset of these quantities are integrated directly over a 3D comoving sphere of radius $\rfhc$, including temperature $T$, bolometric X-ray luminosity $\lx$, the gas mass $M_\mathrm{gas}$, the stellar mass $\mstar$, and the 1D projected velocity dispersion $\sigvtrue$. We consider these as idealised observables without systematic errors and expect that they tightly correlate with the host halo mass \citep{Evrard2008,Mantz2016WtG,Farahi2018,Mulroy2019}, but they are difficult or impossible to know exactly from observation. In contrast, we include several realistic observables calculated directly from the mock catalogues described in Section~\ref{sec:dataset}. These are contaminated with observational systematics and more accurately reflect real measurement conditions. These include the total X-ray photon count in both the whole aperture, $\nphot$, and the central region $\nphotfhc$ of our mock eROSITA images, as well as the richness $\ngal$ and total 1D projected velocity dispersion $\sigv$ as measured in our mock spectroscopic catalogue. $\nphot$ is calculated over every pixel in our wide aperture of $\Rproj\leq 2.3\ \hmpc$, while $\nphotfhc$ only counts pixels within $\Rproj\leq \rfhc$. We construct the $\ngal$ and $\sigv$ probes from the population of galaxies selected within a small dynamical cylinder of aperture $\Rproj \leq\ 1.25 \hmpc$ and velocity cut $|\vlos|<= 2000\ \kms$ and further refined via $3\sigma$ clipping on the $\vlos$ distribution, as a simplistic form of interloper removal. $\sigvtrue$ and $\sigv$ are both calculated with the gapper estimator of velocity dispersion. 

We evaluate the predictive power of these observables using a power-law model for multi-property cluster statistics as presented in \citet{evrard2014model}. Here, we review the framework of this model and refer the reader to the original paper for further details. Consider a set of observable cluster properties $\mathbf{S}$ and their natural logarithmic counterparts $s_i = \ln S_i$, chosen in some given unit definition. Then consider a desired predictor, in this case, the cluster mass $\mfhc$ and its associated logarithm-scaled value $\mu = \ln (\mfhc/M_p)$, normalised to a fixed pivot mass $M_p = 10^{14}\ \hmsun$. In this setting, we can build a model in which the vector of features $\mathbf{s}$ scales linearly with our logarithmic mass $\mu$,
\begin{equation}\label{eqn:scaling}
    \langle \mathbf{s}|\mu\rangle = \boldsymbol{\pi} + \boldsymbol{\alpha}\mu,
\end{equation}
where the vectors $\boldsymbol{\pi}$ and $\boldsymbol{\alpha}$ are the intercepts and slopes, respectively, of the individual scaling laws. These parameters are redshift-dependent, and, in our analysis, we fit distinct $\boldsymbol{\pi}$ and $\boldsymbol{\alpha}$ for each redshift bin of our mock catalogue. We then assume that the uncertainty about this mass-observable relationship is normally distributed with the mean given in Equation~\eqref{eqn:scaling}. \rev{The accuracy of this} assumption is tested and verified against simulation \citep{farahi2018localized,anbajagane2020stellar}.  In this case, the joint distribution of features $\mathbf{s}$ is completely described by the covariance matrix,
\begin{equation}\label{eqn:covariance}
    C_{i j}=\left\langle\left(s_i-\left\langle s_i \mid \mu\right\rangle\right)\left(s_j-\left\langle s_j \mid \mu\right\rangle\right)\right\rangle . 
\end{equation}
We then assume a local, first-order Taylor expansion to the halo mass function \citep{rozo2014comparative},
\begin{equation}\label{eqn:hmf}
    n(\mu) = Ae^{-\beta\mu},
\end{equation}
where $A$ and $\beta$ are the local amplitude and slope of the mass function evaluated at the pivot $\mu=0$. This allows us to invert the relationship of Equation~\eqref{eqn:scaling} to derive the predictive mean and variance of the mass given the information contained in the observables,
\begin{align}\label{eqn:predmean}
    \langle\mu \mid \mathbf{s}\rangle &=\left[\boldsymbol{\alpha}^T \mathbf{C}^{-1}(\mathbf{s}-\boldsymbol{\pi})-\beta\right] \sigma_{\mu \mid \mathbf{s}}^2,\\\label{eqn:predvariance}
    \sigma^2_{\mu|\mathbf{s}} &= \left(\boldsymbol{\alpha}^T\mathbf{C}\boldsymbol{\alpha}\right)^{-1}.
\end{align}
This procedure can determine the minimal scatter on cluster mass of any set of observables $\mathbf{S}$. 

\begin{figure}
    \centering
    \includegraphics[width=\linewidth]{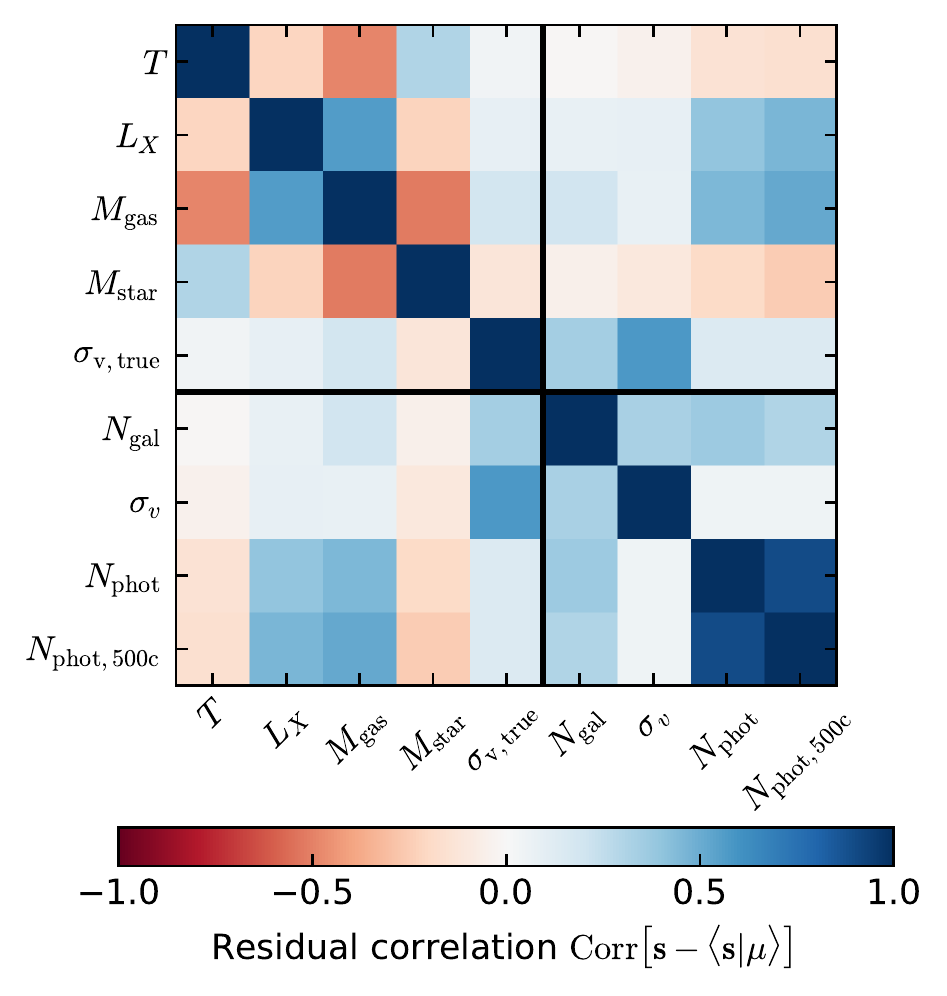}
    \caption{Residual correlation matrix of the scalar observables which form our comparative baseline for cluster mass inference. This covariance is presented under the power law mass dependence of the model described in Section \ref{sec:baseline}. The values shown here are equal to the covariance described in Equation \ref{eqn:covariance} scaled by each observable's individual scatter, i.e., $C_{ij}/(\sigma_{s_i|\mu}\sigma_{s_j|\mu})$. The black vertical and horizontal lines divide the idealized and the realistic observables, as described in Table \ref{tab:observables}. The correlation here is taken from clusters in our mock catalog at redshift $z=0.17$.}
    \label{fig:correlation}
\end{figure}

Using Equation~\eqref{eqn:predvariance}, we measure the predictive variance of each individual observable in our baseline suite and list them in Table~\ref{tab:scatters}. We also perform an exhaustive search over all combinations of ideal and realistic observables to find those with the lowest scatter. The combinations with the least scatter for their number of included observables are also shown in Table~\ref{tab:scatters}. The improvement in these combinations of observables is due to the strong correlations between their mass-dependent residuals, shown as a heatmap in Figure \ref{fig:correlation}.

The scatter we find on observable mass proxies suggests that they are a reliable baseline for mass estimation. Among the idealised observables, $\mgas$ has the lowest scatter at $\sim8.5\%$, consistent with empirical data in \citet{Mulroy2019}. The most direct, yet idealised observable for eROSITA measurements, the bolometric X-ray luminosity $\lx$, only reaches a mass scatter of $\sim 26\%$. Among the realistic observables, $\nphotfhc$ has the lowest scatter at $\sim34\%$ followed by $\ngal$ and $\nphot$ at $\sim40\%$. The improvement in using $\nphotfhc$ over $\nphot$ is due to the removal of AGN contaminates outside of the $\rfhc$ cut. Another notable point is the poor performance of the total velocity dispersion $\sigv$ ($\sim70\%$ scatter) and its idealised counterpart $\sigvtrue$ ($\sim 38\%$ scatter). It is evident both from this study and previous works that dynamics measurements from galaxy spectra are high scatter predictors of cluster mass \citet{saro2013toward}, particularly for mass definitions at low radii. These results suggest that simple linear fits to dynamical measurements could be poor predictors of $\mfhc$, and motivate the use of nonlinear fits such as the ML models in Section \ref{subsec:xraydyn}.

Figure \ref{fig:correlation} shows strong correlations between the mass residuals of various observables, suggesting that we can combine their information into better mass estimators (Table \ref{tab:scatters}). The correlation matrix follows that of \citet{Farahi2019correlation}, with the only discrepancy being in the anti-correlation of $T-\mgas$, which was also found in \citet{kravtsov2006new}. These results show that the most informative two-observable probe of $\mfhc$ is \rev{$\{\mgas, \mstar\}$} with a residual scatter of $\sim 4.9\%$, better than the commonly used probe of $Y_X \propto \mgas T$ \citep{kravtsov2006new} which we measure to have a mass scatter of $\sim 5.1\%$. We find that with only three observables, $T$, $\mgas$, and $\mstar$, the mass scatter can be constrained to the level of $\sim 4\%$. Including all idealised observables, this reduces to $\sim 3.9\%$. However, this idealised scatter is far from what is possible with linear combinations of realistic observables. The combination of all realistic observables (i.e., $\{\ngal,\sigv,\nphot,\nphotfhc\}$) only reaches a minimum scatter of $\sim30\%$, which is still higher than that of the individual idealised observables apart from $\sigvtrue$. These mass scatters of realistic and idealised proxy observables set the lower and upper bounds, respectively, for what we can expect to be the predictive performance of the ML models presented in subsequent sections.

\section{CNNs for Single-Band X-ray}\label{sec:xray_only}
In this section, we describe, implement, and test CNN models for X-ray mass estimation in the context of eROSITA mock observations generated in Section~\ref{sec:dataset}. The models presented in this section take the single-band photon maps as input and produce an estimate of logarithmic $\mfhc$ as output. This problem framework allows for direct comparison to baseline scalar proxies derived from photon counts, as well as previous works in deep learning X-ray modelling presented by \citet{ntampaka2019deep}, \citet{Green2019}, and \citet{yan2020galaxy}.

\subsection{Modelling}\label{subsec:xray_modeling}

The data-driven ML models applied in this work are deep neural networks \citep{bengio2017deep}, a class of popular data science tools that have been taught to model increasingly complex problems, both in science \citep{carleo2019machine} and beyond \citep{lecun2015deep}. Functionally, we can think of deep neural networks as a class of highly non-linear and differentiable functions $f(\mathbf{x};\boldsymbol\theta)$, where $\mathbf{x}$ is the model input (in our case, X-ray images) and $\boldsymbol\theta$ is the model parameterization \rev{(i.e. weights and biases)}. The objective of training a neural network is to find some parameter setting $\boldsymbol\theta^*$ that minimises a loss function $\mathcal{L}(\boldsymbol\theta, \{x, y\})$ given a training sample $\{x, y\}$. Typically, this optimisation is done using a version of gradient descent. The loss function characterises what we want the neural network to do. In our case, we have a dataset $\mathcal{D} = \{(\mathbf{x}^{(i)}, m_i)\}_{i=1}^N$ of $N$ pairs of X-ray images $\mathbf{x}^{(i)}$ and known logarithmic masses $m = \log_{10} \mfhc$. Our goal is to learn from this dataset a model parameterization that best predicts the masses given the X-ray images. To do so, we use a mean squared error (MSE) loss averaged over the training set,
\begin{equation}\label{eqn:loss}
    \mathcal{L}(\boldsymbol\theta) = \sum_{i=1}^N \left[f\left(\mathbf{x}^{(i)}, \boldsymbol\theta\right) - m_i\right]^2.
\end{equation}
 Optimisation of this function in $\boldsymbol\theta$ results in a neural network tuned to predict the mean of the posterior distribution of the logarithmic cluster mass given the X-ray image input \citep{ho2021approximate}.

The specific neural network model used to process X-ray images for this task is a convolutional neural network \citep[CNN;][]{lecun1998gradient}. CNNs are widely recognised as the gold standard for computer vision tasks in machine learning, due to their innate ability to learn and understand non-linear features from subregions of images. We use a feed-forward CNN architecture inspired by the model introduced in \citet{ntampaka2019deep}. Following the $128\times128$ single-channel input image, the network architecture is as follows:

\begin{enumerate}
    \item 2D convolutional layer with 24 filters of size $5\times5$
    \item 2D convolutional layer with 10 filters of size $3\times3$
    \item Max-pooling layer of pooling size $4\times4$
    \item 2D convolution layer with 24 filters of size $5\times5$
    \item 2D convolutional layer with 10 filters of size $3\times3$
    \item Max-pooling layer of pooling size $2\times2$
    \item Flattening operation
    \item Dense layer with output size of $128\times1$
    \item Dense layer with output size of $128\times1$
    \item Dense layer with output size of $64\times1$
    \item Dense layer with output size of $1\times1$
\end{enumerate}
The final layer with one node represents the predictor variable, i.e., the logarithmic $\mfhc$. All layers except the last use a rectified linear activation function (ReLU) and an L2 weight regularisation with magnitude $10^{-4}$. The model is trained using the Adam optimisation scheme \citep{kingma2014adam} at a learning rate of $10^{-3}$.

\rev{We use an eight-fold cross-validation procedure to evaluate the predictive performance of these models on our mock catalogue. To avoid potential train-test data leakage caused by correlated formation environments, the Magneticum clusters are first separated into eight equally-sized quadrants in the simulation volume. Each quadrant is assigned to a unique cross-validation fold. Eight independent models are then each trained on seven folds ($\sim 85\%$ of the data) and tested on the last eighth fold. The tested fold is not repeated among all of our models, so we end with a set of predictions for all clusters in our catalogue. This process ensures that no clusters (or their local neighbors) can be used for both training and testing at the same time. During training, the training folds are further randomly divided into $90\%$ training data and $10\%$ validation data. We use an early-stopping criterion of no validation loss improvement over 20 epochs.} Model training generally converges within $\sim 150$ epochs. During testing, we measure predictive bias and scatter averaged across the \rev{eight} cross-validation folds.

\subsection{Results}

\begin{table}
    \centering
    \begin{tabular}{l|c|c}
        \toprule
        Model & Scatter ($\mathrm{dex}$) & Scatter ($\%$) \\
        \midrule
        ${T}$ & 0.0510 & 11.75 \\
        ${L_X}$ & 0.1144 & 26.33 \\
        ${\mgas}$ & 0.0367 & 8.46 \\
        ${\mstar}$ & 0.0545 & 12.56 \\
        ${\sigvtrue}$ & 0.1631 & 37.56 \\
        \midrule
        ${\ngal}$ & 0.1725 & 39.72 \\
        ${\sigv}$ & 0.3057 & 70.39 \\
        ${\nphot}$ & 0.1738 & 40.02 \\
        ${\nphotfhc}$ & 0.1479 & 34.07 \\
        \midrule
        \rev{${\mgas, \mstar}$} & 0.0214 & 4.93 \\
        \rev{${T, \mgas, \mstar}$} & 0.0176 & 4.06 \\
        \rev{${T, L_X, \mgas, \mstar}$} & 0.0170 & 3.90 \\
        \rev{${T, L_X, \mgas, \mstar, \sigvtrue}$} & 0.0169 & 3.90 \\
        \midrule
        \rev{${\ngal, \nphotfhc}$} & 0.1252 & 28.82 \\
        \rev{${\ngal, \sigv, \nphotfhc}$} & 0.1198 & 27.59 \\
        \rev{${\ngal, \sigv, \nphot, \nphotfhc}$} & 0.1187 & 27.34 \\
        \midrule
        \rev{Single-band X-ray CNN} & \rev{0.0773} & \rev{17.80} \\
        \rev{Multi-band X-ray CNN} & \rev{0.0703} & \rev{16.18} \\
        \rev{X-ray+Spec-z CNN} & \rev{0.0691} & \rev{15.90} \\
        \bottomrule
    \end{tabular}
    \caption{Residual scatter for predicting $\mfhc$ with baseline proxy models (Section \ref{sec:baseline}) and machine learning models (Sections \ref{sec:xray_only} and \ref{sec:multiwavelength}). Each row shows the predictive scatter in $\mathrm{dex}$ and percent, for reference. Models are divided by their type, in the order: idealised observables, realistic observables, combinations of ideal proxies, combinations of realistic proxies, and machine learning models. We performed an exhaustive search over all combinations of observables, but only listed here the lowest scatter combination for each number of observables. The lowest scatter in each section is highlighted, for reference. The reported scatter for each observable was calculated using Equation~\eqref{eqn:predvariance}.}
    \label{tab:scatters}
\end{table}

\begin{figure*}
    \centering
    \includegraphics[width=\linewidth]{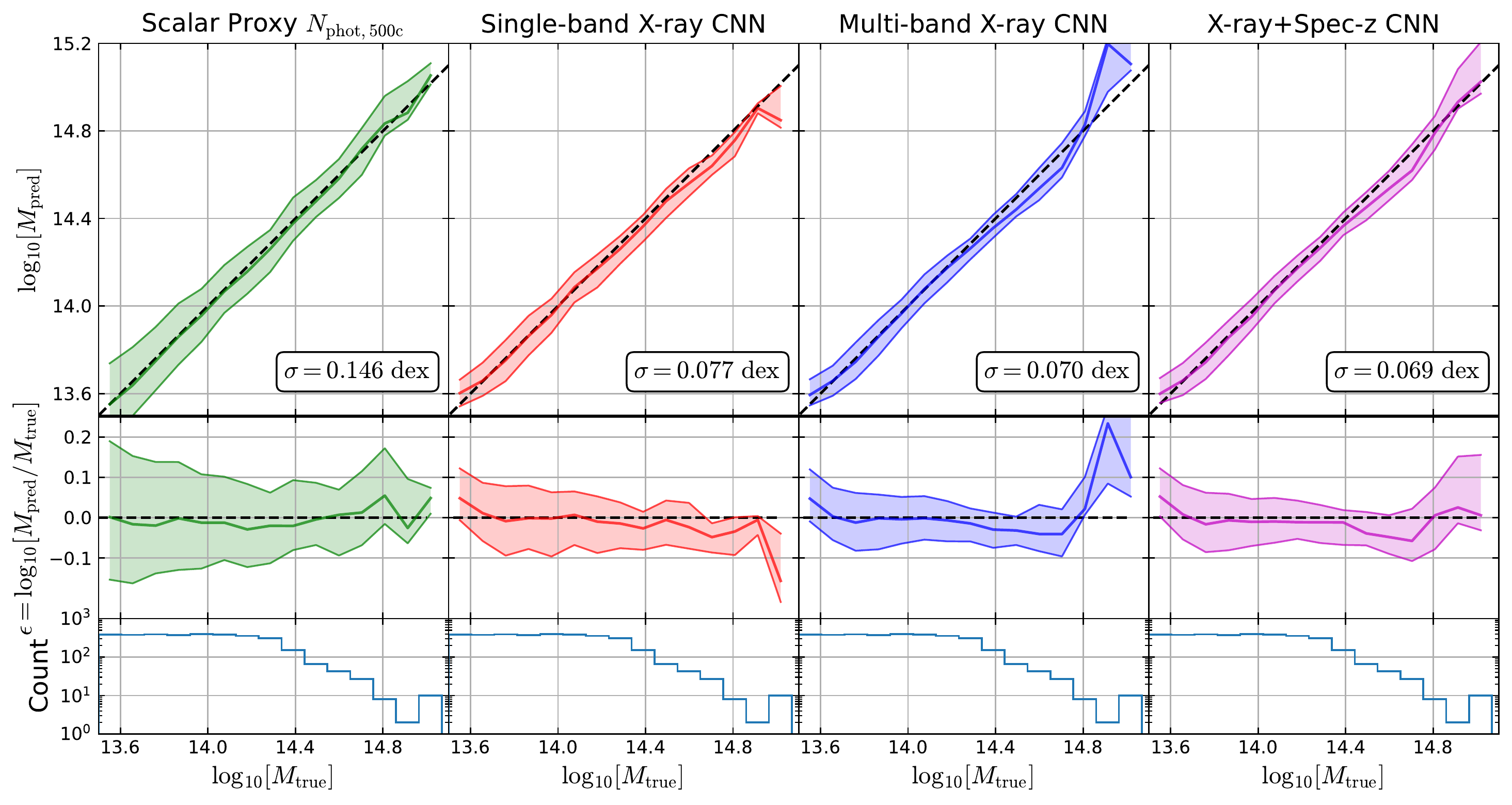}
    \caption{True versus predicted cluster mass for baseline scalar proxies and ML models. Each column displays the predictive performance of a single model in our study. The upper subplots show the median and $[16^{\rm th}$-$84^{\rm th}]$ percentile confidence intervals of the predicted masses, while the \rev{middle} subplots show the corresponding residuals (Equation \ref{eqn:residuals}). \rev{The bottom plots show the histogram of true cluster masses in the test set, which is the same for each model.} From left-to-right, these models are the proxy model for the number of photons detected within $\rfhc$ (Section \ref{sec:baseline}), the single-band X-ray CNN model (Section \ref{sec:xray_only}), the multi-band X-ray CNN model (Section \ref{subsec:multiband}), and the CNN model using both X-ray and dynamical information (Section \ref{subsec:xraydyn}), at maximum galaxy sampling. The mass definition used here for $\mtrue$ and $\mpred$ is $\mfhc$ with units $\hmsun$.}
    \label{fig:truepred}
\end{figure*}

Figure \ref{fig:truepred} shows the distribution of predicted masses and residual scatter of the clusters in our independent test set relative to the best realistic observable, $\nphotfhc$. We characterise the performance of this model in terms of predictive residuals, defined as
\begin{equation}\label{eqn:residuals}
    \epsilon = \log_{10}\left[\frac{M_\mathrm{pred}}{\mfhc}\right].
\end{equation}
The variance of this quantity integrated over our test set asymptotically approaches the Bayesian predictive variance of our estimators under a fixed-variance Gaussian likelihood and is complementary to the predictive variances described in Section \ref{sec:baseline}. \rev{We note that the $\nphotfhc$ scatter shown in Figure \ref{fig:truepred} was calculated by fitting a power-law regression on the training set and evaluating it on the test set. It is slightly different to the variance estimate calculated with Equation~\eqref{eqn:predvariance} due to the finite size and specific prior of the test set.}

The mass predictions of our CNN model demonstrate remarkably low scatter, with an average residual of \rev{$\sim 0.1\%$} and a residual scatter of $\sim 18\%$. This scatter is lower than the scatter of $\nphotfhc$ by \rev{$48\%$} and even outperforms the maximal combination of realistic observables (i.e. $\{\ngal,\sigv,\nphot,\nphotfhc\}$) by \rev{$\sim35\%$} (see Table \ref{tab:scatters}). \rev{We note that this is also lower than the scatter of real X-ray measurements calibrated with weak lensing, which is on the order of $20\%-50\%$ \citep{zhang2008locuss, mahdavi2013joint}.} Our models are also \rev{$\sim 32\%$} more accurate than the idealised $\lx$ proxy model, suggesting that the CNN finds additional information in the X-ray photon maps beyond the bolometric luminosity. An example of such information can be the surface density profile, which is informative of $\mgas$, a much lower scatter mass estimator. Another piece of information can be morphology measures, including concentration. Morphology measures are correlated with the formation or accretion that can explain scatter about the mass--X-ray bolometric luminosity relation \citep{Hartley2008,Parekh2015,Fujita2018,Farahi2020aging_halos}. 

\citet{ntampaka2019deep} and \cite{yan2020galaxy} used neural network estimators trained on mock X-ray observations built from the IllustrisTNG \citep{nelson2019illustristng} and BAHAMAS \citep{mccarthy2016bahamas} hydrodynamical simulations, respectively. Both studies built mock catalogues from models of ICM emission and configured observation time and resolution for the Chandra telescope. Relative to this previous work, the mocks used in this study have a lower signal-to-noise ratio, with eROSITA's lower resolution and observation time, and more realistic observation systematics, with the newly added inclusion of instrument response and AGN contamination. As a result, the residual mass scatters reported in these studies, $\sim12\%$ and $\sim16\%$, respectively, are inevitably lower than the scatter derived from our single-band models ($\sim 18\%$). This impact of adding realistic errors to mock data has been seen in other works as well. For example the SZ-based estimators of \citet{de2022deep} saw a similar, notable increase in scatter between mass predictions on idealised ($12\%$ scatter) and Planck-like ($20\%$ scatter) mock observations. We argue that the constraints presented here will serve as accurate and robust forecasts for the true bounds of the mass reconstruction of clusters in real eROSITA data.

\citet{Green2019} produced a similar study using ensemble regressors trained on Chandra and eROSITA mocks from the Magneticum simulation. This model extracted morphological parameters such as surface brightness concentration, smoothness, and asymmetry from each X-ray image and used them as features in a random forest regression \citep{breiman2001random}. These models achieved a $\sim 16\%$ mass scatter for estimators when trained on either Chandra or eROSITA X-ray maps, a notable result considering the differences in signal-to-noise and telescope design of each sample. This is a $\sim 2\%$ improvement on the single-band mass scatter achieved in this work. However, compared to this work, the mocks used in \citet{Green2019} did not include the presence of AGN sources and were calculated over a shallow redshift range $0\leq z \leq 0.29$, compared to the range $0\leq z\leq 0.47$ of this analysis. In addition, the calculation of some morphological parameters in \citet{Green2019} requires assumed knowledge of the cluster $\rfhc$, which in practice is subject to scatter.  In contrast, the mass constraints presented in this work implicitly incorporate uncertainties of AGN contamination, redshift-dependent effects, and imperfect knowledge of $\rfhc$, and therefore are a high-fidelity forecast of the expected eROSITA mass reconstruction. Nonetheless, we recommend that these methods, and those of \citet{ntampaka2019deep} and \cite{yan2020galaxy}, be evaluated in equal contexts in a future comparative study.

\section{Multiwavelength Models}\label{sec:multiwavelength}
One challenge of next-generation survey science will be how to usefully combine nonlinear information in multiwavelength probes to perform increasingly precise astronomical inference. 
In this section, we study improvements to the single-band CNNs of the previous section by including multi-band and spectroscopic information in our model input.

\subsection{Multi-band X-ray}\label{subsec:multiband}
The stratification of X-ray photons into several wide energy bands is useful for constraining cluster mass \citep{yan2020galaxy}. This stratification allows one to utilise the different spectral profiles of ICM, AGN, and background noise sources to better characterise the physical modelling. eROSITA's instrument will scan the sky in the X-ray band, recording both photon counts and pulse height amplitudes (PHAs), a rough proxy for photon frequency. This will then allow us to roughly split our X-ray observations into different photon energy bands, namely the soft ($0.5-1.2\ \mathrm{keV}$), medium ($1.2-2.0\ \mathrm{keV}$), and hard bands ($2.0-7.0\ \mathrm{keV}$).

Here, we apply a neural network on the multi-band X-ray images of Section~\ref{subsec:data_xray} to place tighter constraints on cluster mass. The modelling of multi-band X-ray images is nearly identical to that of single-band images in Section \ref{subsec:xray_modeling}, except for the fact that we are now dealing with multi-channel inputs. The photon maps, now split into eROSITA's soft, medium, and hard bands, are concatenated together like RGB channels in a photograph. They are then subject to the same CNN architecture as detailed for the single-band images, except that the convolutional filters in the first layer have three channels instead of one. The multi-band models are trained using exactly the same procedure as the single-band models, including the same optimiser, learning rate, and ten-fold cross-validation split.

Figure~\ref{fig:truepred} shows the distribution of mass predictions made by our multi-band model on the independent test set. Remarkably, the use of multi-band images reaches a scatter of \rev{$\sim16.2\%$}, a reduction of \rev{$9\%$} beyond the single-band scatter. This suggests that the model should utilise the knowledge of band separation to form its predictions better. This interpretation is further explored in Section~\ref{subsec:bandatt}. The residual scatter of the multi-band model greatly outperforms both the idealised $\lx$ and realistic observables. \rev{Also, like the single-band models, the multi-band models have a low mean residual of $\sim-0.2\%$ but exhibit mean biases at the high-mass end. This is a result of limited training and test data in the high-mass regime and not necessarily indicative of poor modelling, as evident in previous works \citep{ntampaka2019deep, Green2019}.}

\rev{During the editorial review of this paper, the preprint of \citet{krippendorf2023erosita} was made public wherein the authors applied a very similar CNN architecture to mock multiband eROSITA images from the eFEDS simulations \citep{comparat2019active}. In this approach, the authors split eROSITA observations into ten energy bands, which served as distinct channels for input into their CNN model. They then fit a regression over cluster mass using a maximum likelihood loss function. Due to methodological differences between \citet{krippendorf2023erosita} and this work (e.g.~the type of base simulation, method of mock generation, cluster selection function, image pre-processing, and training loss), it is not possible to make a thorough, apples-to-apples comparison of the two approaches.  Despite these differences, it’s worth noting that the conclusions of both manuscripts largely agree on the expected level of mass scatter for eROSITA-observed clusters (16.2\% in this work vs 18.8\% from \citet{krippendorf2023erosita}).}

\subsection{Joint X-ray and Spectroscopic Data}\label{subsec:xraydyn}
\begin{figure}
    \centering
    \includegraphics[width=\linewidth]{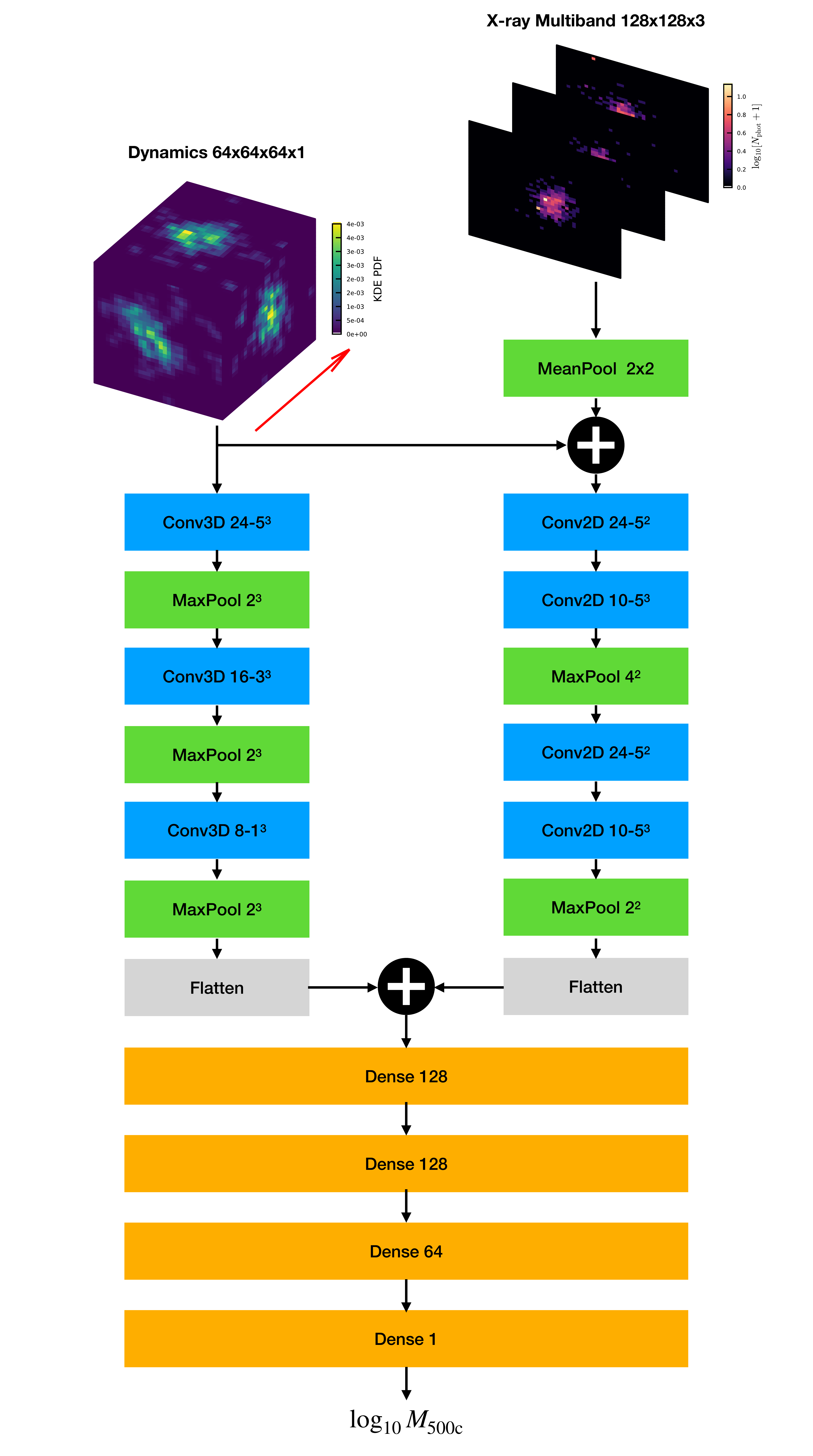}
    \caption{Flowchart neural network architecture for combining X-ray and spectroscopic information as described in Section \ref{subsec:xraydyn}. Each box represents a tensor transformation applied to the previous step. Convolutional layers shown in blue include the number of filters and the filter size. Pooling layers in green show the pooling size. Dense layers in orange show the hidden layer width, and are each followed by a dropout layer of probability $p=0.1$. `Plus' sign operators indicate a concatenation operation. All layers use a Rectified Linear Unit (ReLU) activation function.}
    \label{fig:arch_xraydyn}
\end{figure}

The information encoded in the dynamics of cluster galaxies can be a very potent probe of the system mass. Through the use of spectroscopic follow-up campaigns such as The SPectroscopic IDentification of eROSITA Sources \citep[SPIDERS;][]{furnell2018exploring}, the eROSITA survey will begin to build wide-field multi-wavelength representations of galaxy clusters including both X-ray and dynamical data. Previous works \citep{ho2019robust, ho2021approximate, kodi2020dynamical, kodi2021simulation} have shown that neural networks can effectively model the connection between dynamical information and cluster mass. However, most of these studies have been confined to predictions of larger mass definitions, such as $\mthc$. Part of the reason for this is that dynamics are a very weak probe of cluster core physics, as is evident from the $\sim 38\%$ scatter of $\sigvtrue$ in our baseline analysis (Section~\ref{sec:baseline}). However, it is possible that the inclusion of spectroscopic catalogues will help explain the X-ray systematics, particularly emission from halo environments or projected AGN. So, the question remains as to whether the inclusion of spectroscopic probes will be useful in mass reconstruction for eROSITA clusters or whether X-ray maps capture most of the relevant information.

To test this hypothesis, we construct a neural network architecture to process the X-ray multi-band images from Section~\ref{subsec:data_xray} and dynamical cubes built in Section~\ref{subsec:data_spec}. The architecture is designed to learn informative, localised features from both X-ray and dynamical data and combine them to produce a maximally informative mass reconstruction. The model architecture is split into three sections, a 2D convolutional feature extractor for X-ray-only input, a 3D convolutional extractor for X-ray and dynamical input, and a dense network to take the joint features and produce a cluster mass estimate. The motivation behind this choice is that the X-ray data is the primary driver of the $\mfhc$ information, and the dynamical input is only added to augment or explain the features of the X-ray. Therefore, X-ray inputs have their own path to compression, while the dynamical inputs are inherently tied to the projected X-ray photon counts at their respective sky positions. The X-ray feature extractor has the same design as the convolutional stages of our multi-band model, whereas the dynamical extractor utilises a 3D convolutional architecture following from \citet{kodi2021simulation}. The entire neural network pipeline, including both extractor and dense networks, is trained simultaneously. A representation of the full architecture is shown in Figure~\ref{fig:arch_xraydyn}.

The optimiser, learning rate, and training procedures for this model are identical to those of the single-band model (Section~\ref{sec:xray_only}).
However, because this model utilises 3D convolutions, the forward pass and backpropagation operations are considerably slower to compute. In total, training the 3D convolutional model takes $\sim15\times$ longer than training the 2D convolutional models on a single Nvidia Ampere A100 GPU.

Figure~\ref{fig:truepred} shows that the inclusion of dynamical information adds little improvement to the performance of the multi-band models. The joint X-ray multi-band and dynamical model reaches a test scatter of \rev{$\sim15.9\%$}, which is only slightly smaller than that of the multi-band model at (\rev{$\sim16.2\%$}). These results are even assuming full \rev{spectroscopic} follow-up, i.e., that all galaxies within our selection cut were measured spectroscopically as well. In practice, we only will receive a partial picture of the spectroscopic redshifts of galaxies around each cluster, suggesting that the scatter we achieve here is a lower bound for reality. Although it is possible that the modelling and training applied here are not sufficiently flexible to learn complex multiwavelength relationships, it is rather more likely that most of the information about $\mfhc$ contained in dynamical probes is equally well covered by X-ray measurements \citep[e.g.][]{nagai2007testing}. 


\section{Investigating Cluster Mass Estimates}\label{sec:discussion}
In this section, we analyse the models presented in the previous sections to derive an understanding of the learned behaviour which allows for improved mass estimates. We test several hypotheses presented in the literature to explain improved neural network performance, as well as present new findings from our explainability study.

\begin{figure}
    \centering
    \includegraphics[width=\linewidth]{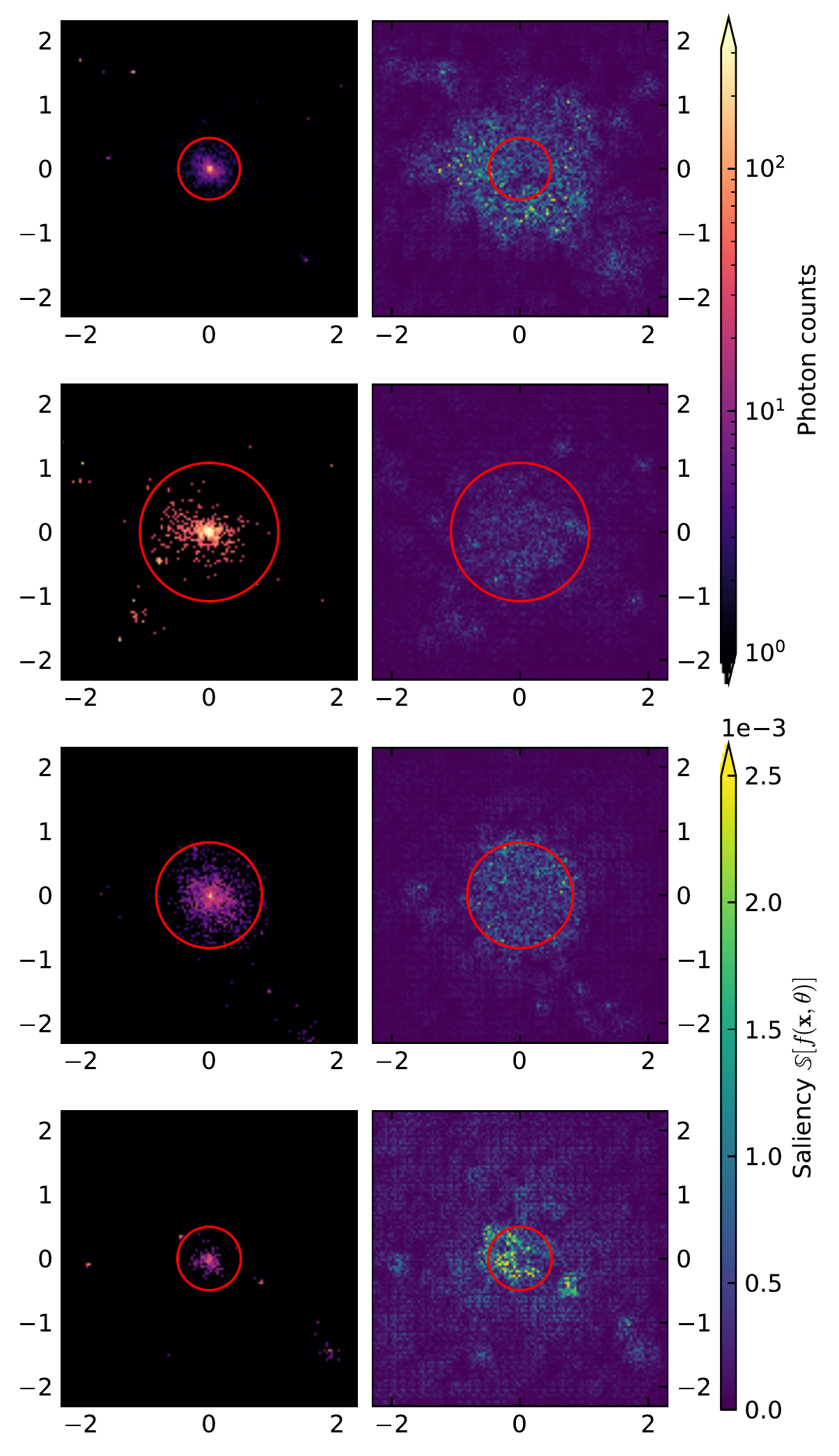}
    \caption{Four example single-band eROSITA X-ray mocks and their corresponding saliency maps when subject to CNN mass estimators. Each row represents a random cluster in our test set. The left column shows the original single-band X-ray image input, while the right column shows the saliency following Equation~\eqref{eqn:saliency}. Each subplot also shows a red circle indicating the $\rfhc$ of each cluster for scale reference. Subplot x- and y-axes are labelled in units of $\hmpc$.}
    \label{fig:ex_saliency}
\end{figure}

The primary diagnostic tool used for our interpretation of neural network behaviour is the saliency map. Saliency maps are popular explainers for image-recognition tasks and function by quantifying the importance of individual pixels in an input image for a given modeling task. The specific method we use is a gradient-based saliency map \citep{simonyan2013deep}, which measures the pixel-wise sensitivity to model outputs by taking the gradient of the output with respect to the input. Specifically, we can consider a saliency map as the output of applying the operator $\mathbb{S}$ to a given model output $f(\mathbf{x};\boldsymbol\theta)$ evaluated at a single input image $\mathbf{x}$, as defined:
\begin{equation}\label{eqn:saliency}
    \mathbb{S}\left[f(\mathbf{x};\boldsymbol\theta)\right] = \left|\left.\frac{\partial f(\mathbf{x};\boldsymbol\theta)}{\partial \mathbf{x}}\right|_{\mathbf{x}=\mathbf{x}}\right|.
\end{equation}
The result of this operation is an image of the same dimensionality as $\mathbf{x}$, wherein each pixel is a measure of the local sensitivity of the model output with respect to the corresponding input pixel. Here, we use the absolute value of the gradient as a sensitivity metric instead of a measure of relative directional change. In general, the function $f(\mathbf{x};\boldsymbol\theta)$ can be any differentiable parametric function, but, in this work, we consider $f$ to be estimators of logarithmic mass, either as the scaling models from Section \ref{sec:baseline} or the neural network models from Sections \ref{sec:xray_only} and \ref{sec:multiwavelength}.  For our application, a high value derived from Equation~\eqref{eqn:saliency} suggests that the current setting of the pixel value is important for mass prediction. Figure~\ref{fig:ex_saliency} shows several examples of single-band X-ray images and their corresponding saliency maps when subject to our CNN mass estimator. 

As an example, we derive the expected saliency for a baseline scaling model using the $\nphot$ 
observable (Section~\ref{sec:baseline}). Following our proxy formalism \citep{evrard2014model}, a single scalar observable $s_a$ will predict a mean logarithmic mass $\ln M$ as,
\begin{equation}\label{eqn:scalingscalar}
    \left\langle \ln M \mid s_a\right\rangle = (s_a-\pi_a)/\alpha_a - \beta \sigma^2_{\ln M \mid a},
\end{equation}
where $\pi_a$ and $\alpha_a$ are the scaling intercept and slope parameters for proxy $s_a$, $\beta$ is the local slope of the halo mass function (Equation~\eqref{eqn:hmf}), and $\sigma^2_{\ln M \mid a}$ is the proxy predictive variance (Equation~\eqref{eqn:predvariance}). 
For a photon count based proxy, our proxy is $s_a = \ln \nphot = \ln\left[\sum_{ij}n_{ij}\right] = \ln \left[\sum_{ij}\left(10^{x_{ij}}+1\right)\right]$, wherein $n_{ij}$ are the photon counts in each pixel and $x_{ij}$ are their transformed counterparts after normalisation with Equation \ref{eqn:inputscaling}. We note that the estimator for the $\nphotfhc$ proxy follows this same form, except summed over a smaller radius. Using this definition, we can turn our proxy model of mean logarithmic mass (Equation \ref{eqn:scalingscalar}) into a mass estimator,
\begin{equation}
    g(\mathbf{x};\boldsymbol\theta) = \frac{\left\langle \ln M \mid \ln \nphot\right\rangle}{\ln 10},
\end{equation}
where $\boldsymbol\theta = \left\{\pi_{\ln\nphot},\alpha_{\ln\nphot}, \beta, \sigma^2_{\ln M|\ln\nphot}\right\}$ and the $\ln 10$ factor scales the $g(\mathbf{x};\boldsymbol\theta)$ output to $\log_{10}M$. Then, we can apply our saliency operator to this $\nphot$ proxy mass estimator to infer the following closed-form expression:
\begin{equation}\label{eqn:baseline_saliency}
    \mathbb{S}\left[g(\mathbf{x};\boldsymbol\theta)\right]_{ij} = \frac{10^{x_{ij}}}{\alpha_{\ln\nphot}\nphot} \propto \frac{n_{ij}+1}{\nphot}.
\end{equation}
Under this model, the saliency would suggest that the more photons a pixel detects, the more important it is to the model. This is an undesirable property of the $\nphot$ scaling model, especially as we seek to exclude, for example, contaminating AGN sources in exchange for more informative ICM emission. In subsequent sections, we will use this saliency example as a relative comparison to the CNN models under investigation.

\subsection{Cluster cores} \label{subsec:coreatt}

\begin{figure}
    \centering
    \includegraphics[width=\linewidth]{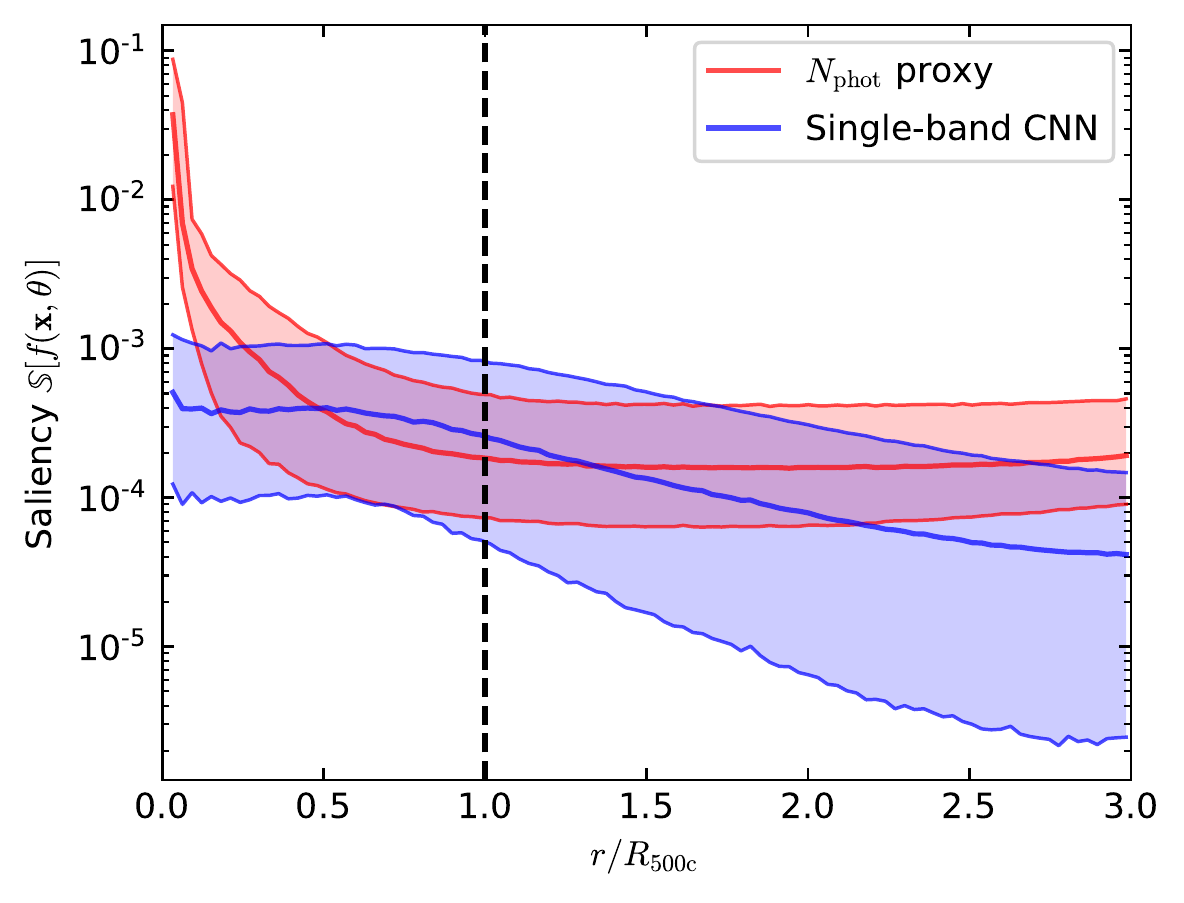}
    \caption{Distribution of input saliencies as a function of radius from the centre of the image for two different mass estimators, a scalar $\nphot$ proxy in red and a CNN model in blue. We show the median and 16-84th confidence interval of each distribution, binned over a radius. Saliency values for the scalar proxy were calculated analytically with Equation~\eqref{eqn:baseline_saliency}, whereas those for the CNN were calculated via backpropagation with Equation~\eqref{eqn:saliency}.}
    \label{fig:core}
\end{figure}

\citet{ntampaka2019deep}, \citet{yan2020galaxy}, \citet{de2022deep}, and \citet{ntampaka2022importance} also apply interpretability methods to explain the decisions of cluster mass estimators. In the former three studies, the authors apply the Google Deep Dream algorithm \citep{mordvintsev2018differentiable} as their interpretability diagnostic, a gradient-based saliency variant that uses gradient ascent to find the perturbation to the input image which would result in the maximal output shift. In the latter study, traditional gradient saliency is used. All four studies qualitatively find that the neural network significantly down-weights the cluster core region when making a mass prediction. The explanation for this behaviour is that the cluster core is unique with its highly stochastic, non-linear physics and thus is a poor signal from which to derive cluster mass. 

Here, we quantitatively test the hypothesis that neural networks learn to down-weight the central regions of an X-ray image when predicting $\mfhc$. Figure~\ref{fig:core} shows the distribution of saliency values assigned to pixels within and outside of the cluster core as a function of radius for our CNN model. The CNN saliency is shown relative to that of the $\nphot$ proxy, the saliency for which we derived analytically in Equation~\eqref{eqn:baseline_saliency}. We compare these two models. In this context, we estimate the relative per-pixel importance given by (i) the $\nphot$ proxy model and (ii) the CNN model.

Figure \ref{fig:core} shows that the radial distribution of saliency values for the $\nphot$ proxy model and the CNN model differ significantly at the central and outer regions of clusters. We observe that, although the neural network models are not totally disqualifying information in the central regions of the cluster, the information in the core is significantly down-weighted considering the number of detected photons in the region. At around $r=\rfhc$, the prediction sensitivity for $\nphot$ and the CNN is close to the same, suggesting that, in this ICM region, the CNN effectively counts the number of photons to predict mass. However, moving below this radius, the $\nphot$ saliency increases exponentially, whereas the CNN saliency remains constant. At a radius of $r=0.07\ \rfhc$, the median $\nphot$ saliency is approximately $10\times$ that of the CNN model. This is caused by the fact that high photon counts in the core of the cluster are extremely crucial to determining the mass prediction with the $\nphot$ proxy, but only mildly important to the neural network. Another interesting result from this diagnostic is that beyond $r>\rfhc$, the CNN starts again to down-weight the importance of pixels. \rev{We note that the CNN model does not directly estimate the true $\rfhc$, though it may infer it from the mass-luminosity relation or the comoving surface brightness profile}. This may be indicative of the fact that CNN recognises that, outside of $r>\rfhc$, the presence of environmental feedback and AGN contamination makes this region an unreliable probe of central cluster mass and therefore disregards it.

\subsection{AGN Attention} \label{subsec:agnatt}

Another hypothesis for improving neural networks is the ability of CNN models to recognise and remove contaminating artefacts within subregions of an image \citep{ho2019robust, kodi2020dynamical}. In the case of X-ray probes, this behaviour would allow the CNN to identify AGN contaminates and mitigate their contribution to the total photon emission. The separation of sources is impossible to do exactly in real observations, but we can study the CNN sensitivity to various sources in the idealised environment of simulation.

\begin{figure}
    \centering
    \includegraphics[width=\linewidth]{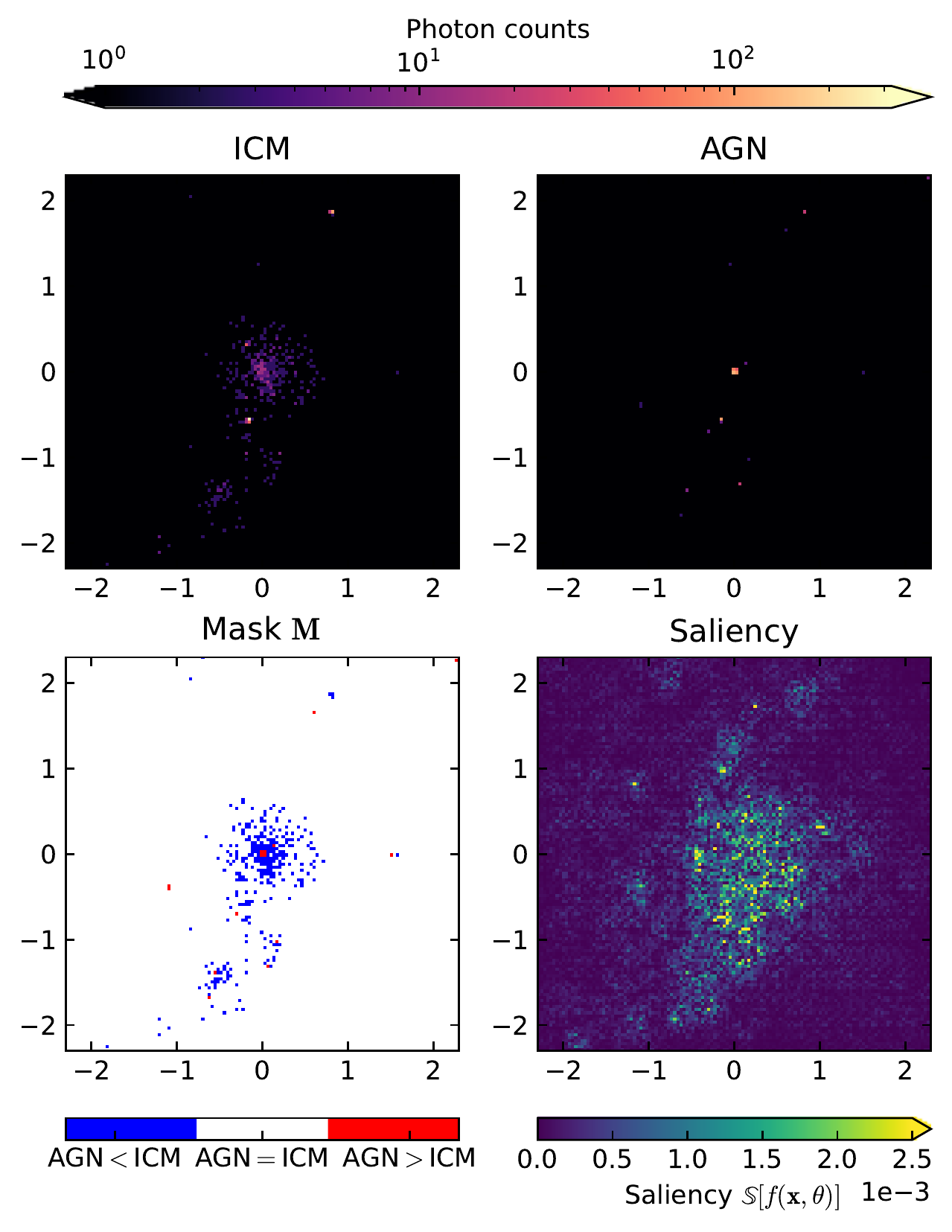}
    \caption{Example ICM and AGN emissions and corresponding saliency for mock X-ray observations of an example cluster. This demonstrates the relative attention paid by our mass estimation model to respective sources in the X-ray image, as discussed in Section \ref{subsec:agnatt}. Top left: ICM emission. Top right: AGN emission. Bottom left: Mask of ICM vs. AGN contributions for use in Equation \ref{eqn:attmetric}. Bottom right: Saliency map for a trained single-band CNN mass estimator derived from Equation \ref{eqn:saliency}.}
    \label{fig:icmagn}
\end{figure}

Inspired by Benchmarking Attribution Methods \citep[BAMs;][]{yang2019benchmarking}, we construct a quantitative test to measure the attention paid by the neural network to ICM and AGN sources. We can consider a given input image $\mathbf{n}$ to consist of a contribution of ICM emission $\mathbf{n}_\mathrm{ICM}$ and AGN emission $\mathbf{n}_\mathrm{AGN}$, such that $\mathbf{n} = \mathbf{n}_\mathrm{ICM} + \mathbf{n}_\mathrm{AGN}$. Then, for each input image $\mathbf{x}^{(i)}$, we can define two binary masks, $\mathbf{M}^{(i)}_\mathrm{ICM}$ and $\mathbf{M}^{(i)}_\mathrm{AGN}$, which quantify which pixels were dominated by each respective source, \rev{i.e., $\mathbf{M}_\mathrm{ICM} = \Theta\left[\mathbf{n}_\mathrm{ICM} - \mathbf{n}_\mathrm{AGN}\right]$ and $\mathbf{M}_\mathrm{AGN} = \Theta\left[\mathbf{n}_\mathrm{AGN} - \mathbf{n}_\mathrm{ICM}\right]$, where $\Theta$ is the element-wise Heaviside function}. Then, we can define a functional $\mathbb{S}_\mathrm{mask}$ which quantifies the average saliency for all pixels integrated over each source mask, as follows:
\begin{equation}\label{eqn:attmetric}
    \mathbb{S}_\mathrm{mask} \left[f(\mathbf{x};\boldsymbol\theta);\mathbf{M}\right] = \frac{1}{\sum_{ij} M_{ij}} \sum_{ij} M_{ij}\cdot \mathbb{S}\left[f(\mathbf{x};\boldsymbol\theta)\right]_{ij},
\end{equation}
where all sums are taken over all matrix elements. Figure \ref{fig:icmagn} shows an example cluster's ICM and AGN components, as well as its binary masks and saliency map. To compare global properties of each source mask, we further derive a summary statistic of Equation \ref{eqn:attmetric} as $\langle \mathbb{S}_\mathrm{mask} \left[f(\mathbf{x};\boldsymbol\theta);\mathbf{M}\right] \rangle$, wherein the operation $\langle\cdot\rangle$ is taken as an average over all inputs $\mathbf{x}^{(i)}$ and corresponding masks $\mathbf{M}^{(i)}$ in our test set.

\rev{For CNN models, we find that the contribution to the saliency from AGN sources is approximately log-normally distributed with median and $[16^{\rm th}$-$84^{\rm th}]$ percentile interval of $\mathbb{S}_\mathrm{mask} \left[f(\mathbf{x};\boldsymbol\theta);\mathbf{M}_\mathrm{AGN}\right] \sim 10^{-3.85_{-0.74}^{+0.71}}$, whereas that of ICM sources is $\mathbb{S}_\mathrm{mask} \left[f(\mathbf{x};\boldsymbol\theta);\mathbf{M}_\mathrm{ICM}\right] \sim 10^{-3.53_{-0.67}^{+0.47}}$. While the range of each distribution is wide, we note that the median value of ICM saliencies is twice that of AGN saliencies.} This is particularly notable given that pixels dominated by AGN emission, that is when $\mathbf{n}_\mathrm{AGN}>\mathbf{n}_\mathrm{ICM}$, receive on average $>13$ times the number of photons as those dominated by ICM emission, a result of the intense radiation emitted from AGN sources. From Equation~\eqref{eqn:baseline_saliency}, this would suggest that AGN-dominated pixels would attend for $13$ times the saliency of ICM pixels for a typical $\nphot$ proxy scaling relation. The fact that the CNN saliency of AGN sources is less than that of ICM sources clearly indicates that CNN models treat AGN sources with reduced importance compared to ICM sources, a definitive departure from the behaviour of scalar proxy-based methods.

However, we note that the AGN attention in our models is nonzero, suggesting that changing AGN photon counts would indeed impact our CNN mass estimates, albeit in a reduced capacity to ICM emission. We propose an explanation for this in that the CNN model must learn the corpus of possible AGN profiles in order to recognise their location and remove their photon counts. In this case, any pixel-wise deviation of the established profile will significantly affect the recognition of the AGN source, thereby deviating the result. Qualitative evidence for this behaviour exists in Figure \ref{fig:icmagn} wherein pixels with no detected photons located between the extended AGN source and the central ICM are given high saliency. If these pixels were to be non-zero, the model would recognise evidence of the extended AGN source as being part of the ICM and accordingly increase its mass estimate. We recommend exploring this hypothesis in future studies. 

\subsection{Multi-band Attention} \label{subsec:bandatt}
\begin{figure}
    \centering
    \includegraphics[width=\linewidth]{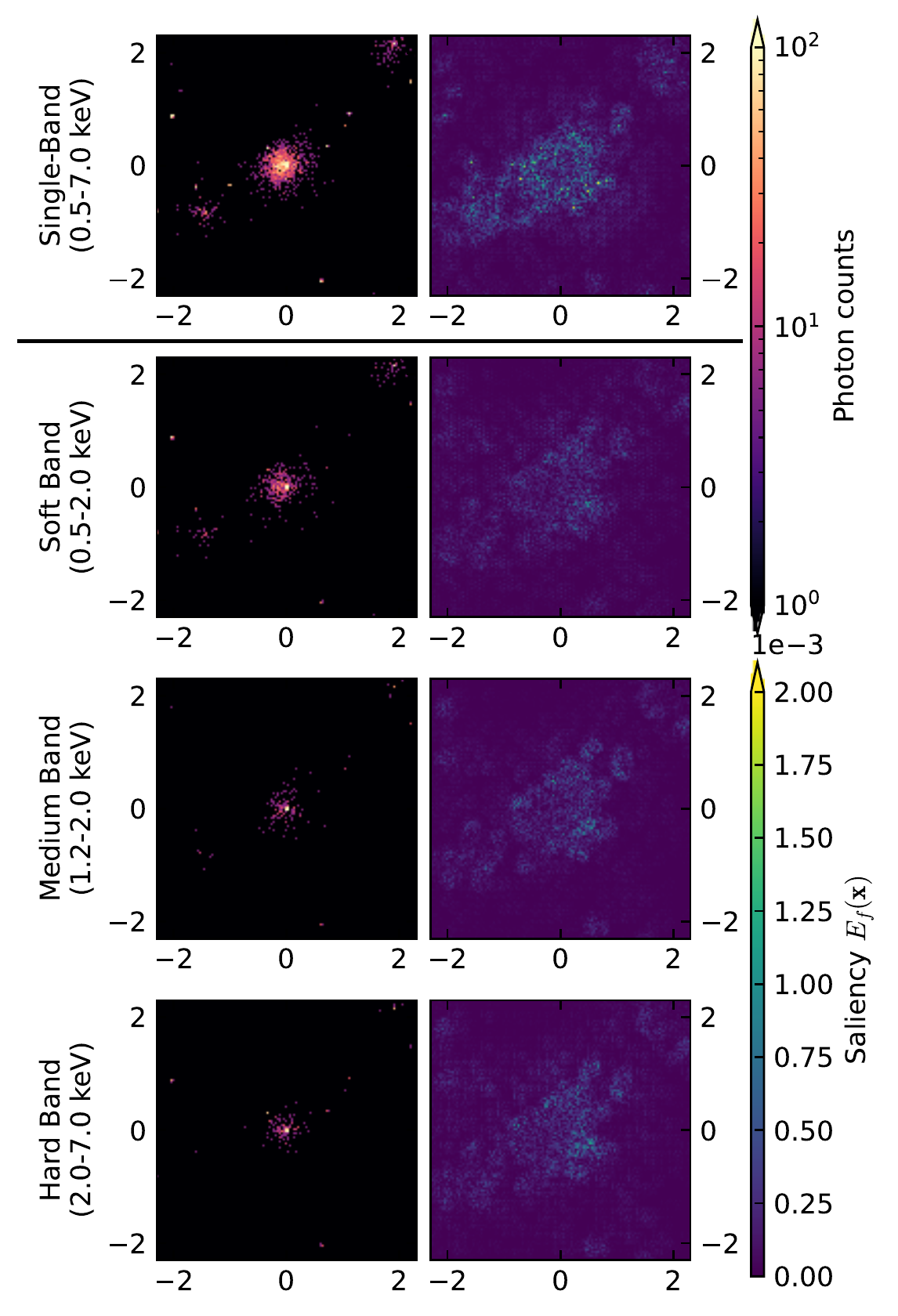}
    \caption{X-ray photon and saliency maps across different X-ray bands for a single example cluster in our test set. The top row shows the input X-ray photon map and resultant saliency map when doing mass inference with a single-band X-ray CNN. The bottom three rows show the photon maps and saliencies for the same mass inference with a multi-band X-ray CNN. The bottom images are separated into soft, medium, and hard bands.}
    \label{fig:attMB}
\end{figure}

In Section \ref{subsec:multiband}, we showed that stratification of X-ray photon maps into soft, medium and hard bands improves the mass prediction performance of the neural network model. The intuition of previous work suggests that peak ICM emission occurs in the soft bands, while AGN emission is more dominant in the hard bands \citep{biffi2018agn}. Therefore, the separation of the soft band from other energies would provide a cleaner sample from which we can study the ICM. We apply the saliency maps to the multi-band models to study the sensitivity of predictions to each energy band. 

Figure~\ref{fig:attMB} shows the saliency maps for an example multi-band CNN prediction compared to the saliency map of a single-band prediction. We derive two  interpretations on the behaviour of multi-band CNN models: \rev{First, the distribution of pixel saliencies in the soft, medium, and hard bands are very nearly identical, with median value and $[16^{\rm th}$-$84^{\rm th}]$ percentile intervals of $10^{-4.15_{-0.68}^{+0.56}}$, $10^{-4.17_{-0.68}^{+0.59}}$, and $10^{-4.15_{-0.72}^{+0.60}}$, respectively.} The saliency intensities follow roughly the same spatial distribution for each cluster image. This is suggestive that multi-band information is closely shared across the channels by the CNN architecture but does not allow us to draw any definitive conclusions about the relative importance of measuring one band over another. Second, we clearly notice that AGN emission present in the input X-ray photon maps is not as significant in the saliency of the multi-band models as they are for the single-band models. \rev{This is reflected in repeating the analysis of Section~\ref{subsec:agnatt} for the multi-band models, in which we find that the saliency of AGN-dominated pixels is $10^{-3.91_{-0.74}^{+0.61}}$
versus that of ICM-dominated pixels $10^{-3.77_{-0.63}^{+0.49}}$.} We can then gather that the AGN-to-ICM saliency ratio for multi-band models is 50\% lower than that of the single-band models. We suggest that this is a result of the spectral separation of AGN and ICM sources in multi-band images.


\section{Conclusion}\label{sec:conclusion}

In this work, we present forecasts for the expected limits of mass reconstruction for eROSITA clusters with deep learning models. To ensure we place reliable constraints, we validate our methodology on a highly realistic catalogue of mock X-ray observations derived from the Magneticum hydrodynamic simulations. The mock catalog used here includes systematics such as AGN contaminants, cluster morphology, background emission, and eROSITA-specific instrument response, making it the most robust mock X-ray catalogue currently applied to deep learning mass estimation. Using this catalogue, we build a quantitative baseline of scalar mass proxies and characterise their intrinsic scatter and mutual information. We then introduce a CNN architecture based on the one used in \citet{ntampaka2019deep} designed to learn the mapping between photon maps and logarithmic mass $\log_{10} \mfhc$. For single-band, bolometric X-ray photon maps, we can constrain $\mfhc$ masses to within $18\%$ scatter, a factor of two improvements on realistic mass proxies, and a \rev{$32\%$} improvement on an idealised bolometric luminosity $\lx$ measurement. This result suggests that neural networks successfully utilise high-order features to reduce predictive uncertainty.

We go on to suggest that these models can be improved with the inclusion of multiwavelength inputs. We show that a CNN model trained on X-ray photon maps separated into soft, medium, and high energy bands further reduces mass scatter to \rev{$16.2\%$, a $9\%$ reduction from the single-band models}. We also find that the inclusion of dynamical information from spectroscopic follow-up offers little-to-no scatter improvement for $\mfhc$. We argue that $\mfhc$ information contained in cluster dynamics is mostly or entirely covered by photon maps.

Lastly, we investigate a series of hypotheses as to why CNNs show such substantial improvements over conventional scalar observables. We quantitatively demonstrate that CNNs down-weight the importance of photons emitted by cluster cores by greater than a factor of 10, motivating the belief that these features are too noisy for reliable mass inference. We also show that the CNNs significantly downweight, but do not ignore, contamination from AGN emission in X-ray images. Lastly, we perform tests of feature attribution for multi-band X-ray models, but do not find definitive evidence that different X-ray bands contribute more or less to the overall mass prediction. However, we find that multi-band models are better at reducing the importance of AGN emission photons. We suggest this results from better source separation in different photometric bins.

In conclusion, DL models can learn a reliable, informative model of X-ray clusters, infer low-scatter estimates of their mass, and be well-calibrated with realistic mock simulations. Their improvements can be attributed to physically-motivated manipulation of information, including core excision and automatic removal of AGN contamination. We recommend these techniques for further application to the upcoming eROSITA cluster catalogue, complementing existing proxy-based mass estimators.


\section*{Acknowledgements}
\rev{We thank the anonymous referee for the helpful comments and suggestions used to improve the manuscript during the review process.} We acknowledge Klaus Dolag and Antonio Ragagnin for providing the Magneticum 2b simulations used in this work.  MH is supported by the Simons Collaboration on
``Learning the Universe" and was supported by NSF AI Institute: Physics of the Future, NSF PHY-2020295, and the McWilliams-PSC Seed Grant Program. JS, DN, and MN acknowledge support from the NASA ATP Grant (80NSSC22K0821). DN is also supported by NSF (AST-2206055) and NASA (TM3-24007X) grants. \rev{AE is supported by NASA grant (80NSSC22K0476).} The computing resources
necessary to complete this analysis were provided by the Pittsburgh Supercomputing Center. The X-ray mock data set was collected using computational resources at the Maryland Advanced Research Computing Center (MARCC).

\section*{Data Availability}
The Magneticum simulations and respective halo catalogues and PHOX X-ray mocks used in this analysis are publicly available through the Cosmology Web Portal \footnote{\url{http://www.magneticum.org/vos.html}}. Mock eROSITA photometry can be generated using preset configurations in the SIXTE code\footnote{\url{https://www.sternwarte.uni-erlangen.de/sixte/}} \citep{dauser2019sixte}. Pre-generated dynamical catalogues for Magneticum halos are available from the authors upon reasonable request.

All code for the data analysis and machine learning models investigated in this analysis has been made available on Github\footnote{\url{https://github.com/McWilliamsCenter/halo_cnn}}. Pre-processed training and test catalogues and pre-trained models will be made available upon reasonable request.




\bibliographystyle{mnras}
\bibliography{references} 


\bsp	
\label{lastpage}
\end{document}